\documentclass[prl,twocolumn,showpacs,amsmath,amssymb]{revtex4-1}

\usepackage{graphicx}% Include figure files
\usepackage{dcolumn}% Align table columns on decimal point
\usepackage{bm}% bold math
\def\3he{$^3$He}
\def\4he{$^4$He}

\begin{document}

%\preprint{APS/123-QED}

\title{Observation of thermo-mechanical equilibration in the presence of a solid \4he conduit}% Force line breaks with \\

\author{M.W. Ray}
\author{R.B. Hallock}%
\affiliation{%
Laboratory for Low Temperature Physics, Department of Physics,\\
University of Massachusetts, Amherst, MA 01003
}%

\date{\today}% It is always \today, today,

\begin{abstract}
We observe a thermo-mechanical effect when a chemical potential difference is created by a temperature difference imposed between two liquid reservoirs connected to each other through Vycor rods in series with solid hcp \4he.  By creating a temperature difference, $\Delta T$, between the two reservoirs, we induce a rate-limited growth of a pressure difference between the two reservoirs, $\Delta P$. In equilibrium $\Delta P {\it vs.} \Delta T$ is in quantitative agreement with the thermo-mechanical effect in superfluid helium. These observations confirm that below $\sim$ 600 mK a flux-limited flow exists through the solid helium.
\end{abstract}

\maketitle

Kim and Chan\cite{Kim2004a,Kim2004b}, motivated by the work of Ho, Bindloss and Goodkind \cite{Ho1997}, observed a significant shift in the resonant period of a torsional oscillator filled with solid \4he in the vicinity of 100 mK. This shift was interpreted as due to mass decoupling from the oscillator and it was suggested that this was likely evidence for a supersolid phase of solid \4he that was first predicted many years earlier \cite{Penrose1956,Andreev1969,Chester1970,Leggett1970}.  This interpretation of the presence of non-classical rotational inertia (NCRI) continues to be controversial, but has spawned considerable experimental and theoretical activity\cite{Balibar2008}.  Efforts to cause the flow of solid helium through confined geometries by directly squeezing the solid lattice have not been successful\cite{Greywall1977,Day2005,Day2006,Rittner2009}.  In an effort to create a DC flux of atoms through the solid, we took a different approach and by injecting atoms from the superfluid have demonstrated mass transport through the solid\cite{Ray2008a,Ray2009b,Ray2010a} at temperatures below $\sim$ 600 mK. In the present work we create a chemical potential difference between two reservoirs by imposing a temperature difference and observe a rate-limited change of the thermo-mechanical pressure, which indicates that a pathway percolates the solid.

\begin{figure}[b]
\resizebox{2 in}{!}{
\includegraphics{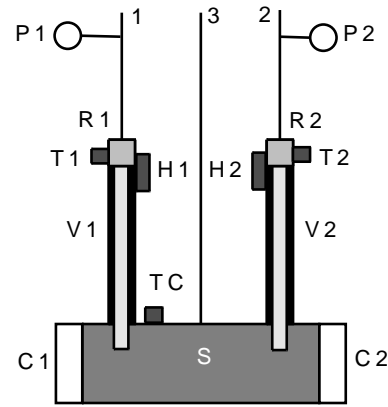}}
\caption{\label{fig:cell} Schematic diagram of the apparatus used to search for flow through solid \4he. Superfluid helium in Vycor allows for an interface between the solid and superfluid at pressures greater than the bulk melting pressure of solid helium.}
\end{figure}

Figure \ref{fig:cell} shows the apparatus used for this work, which is similar to that used previously\cite{Ray2008a,Ray2009b,Ray2010a}.  It consists of a cylindrical copper chamber, S, pierced by two new 3.0 mm dia. epoxy-coated Vycor rods, V1 and V2, which extend to the horizontal axis of chamber S.  The flat ends of V1 and V2 in the cell were coated with epoxy.  With solid \4he in region S, and the pressure in the Vycor less than 37 bar, there is bulk solid \4he in S and superfluid liquid helium in the Vycor rods\cite{Beamish1983,Lie-zhao1986,Adams1987}.  Atop the Vycor rods are thermometers T1 and T2 and liquid reservoirs R1 and R2 that are fed by capillaries (1 and 2 in fig.~\ref{fig:cell}). The temperatures in the reservoirs are controlled using heaters, H1 and H2, so that the helium in R1 and R2 remains a liquid.  A temperature gradient is present across the Vycor; the cell remains at a low temperature. The pressure of the solid is measured by capacitance strain gages \cite{Straty1969} C1 and C2 and the pressures of the reservoirs are measured by pressure gauges P1 and P2 located outside the cryostat.  Using our original apparatus\cite{Ray2008a,Ray2009b}, a chemical potential difference, $\Delta \mu$, could be imposed between the Vycor rods by the creation of a pressure difference $\Delta P = P2 - P1 \neq$ 0 by adding or subtracting atoms from R1 or R2 using lines 1 and 2.  For $TC <$ 600 mK an imposed abrupt increase in pressure $P1$ resulted in an increase in $P2$ accompanied by a rise in the pressure $C1$ and $C2$ with $dP2/dt \approx$ constant, independent of $P2 - P1$.

If one has a superleak in a conducting pathway between R1 and R2, it should be possible to induce a pressure difference $P1 - P2$ by the imposition of a temperature difference $T1 - T2$. In liquid helium, this is the well known fountain effect (or thermo-mechanical effect)\cite{wilks}.  Quantitatively, the fountain effect relates a temperature difference, $\Delta T = T_b - T_a$, between two liquid reservoirs connected by a superleak to a fountain pressure difference, $\Delta P_f$, by the relation
\begin{equation}
\Delta P_f = \int_{T_a}^{T_b}\rho S dT,
\label{eq:fountain}
\end{equation}
where $\rho$ is the liquid density and $S$ is the entropy.

The apparatus shown in fig.~\ref{fig:cell} can be used to utilize the fountain effect in the presence of solid helium.  With solid helium in region S, we use H1 or H2 to vary $T1$ or $T2$, create a chemical potential difference between the reservoirs, R1 and R2,  and then measure the resulting pressures $P1$ and $P2$ in the reservoirs.  If a conducting pathway through the solid is present, we should be able to observe a pressure-temperature relationship given by eq.~\ref{eq:fountain}.  With liquid helium in the cell at $P \approx$ 24 bar, $TC$ = 250 mK, and $T1 \neq T2$ near 1.6 K we confirm that this fountain pressure can be created in quantitative agreement with eq.~1.  In the present report we show that this is also true with solid \4he in the cell.

All of our solid helium samples were grown in the hcp region of the phase diagram from the superfluid at constant temperature using commercial grade \4he ($\sim$300 ppb \3he concentration).  To fill the cell initially, the helium is introduced through line 3 and does not pass through the Vycor rods.  To grow the solid at constant temperature, we begin with the pressure in region S below the bulk melting curve for \4he at low temperature and then add atoms to the solid through lines 1 and 2, which lead to the Vycor.  Because these fill lines remain open we can continue to add atoms to the solid at pressures greater than the bulk freezing pressure of the helium, and grow our solid sample\cite{Ray2010a}.

\begin{figure}
\resizebox{3.5 in}{!}{
\includegraphics{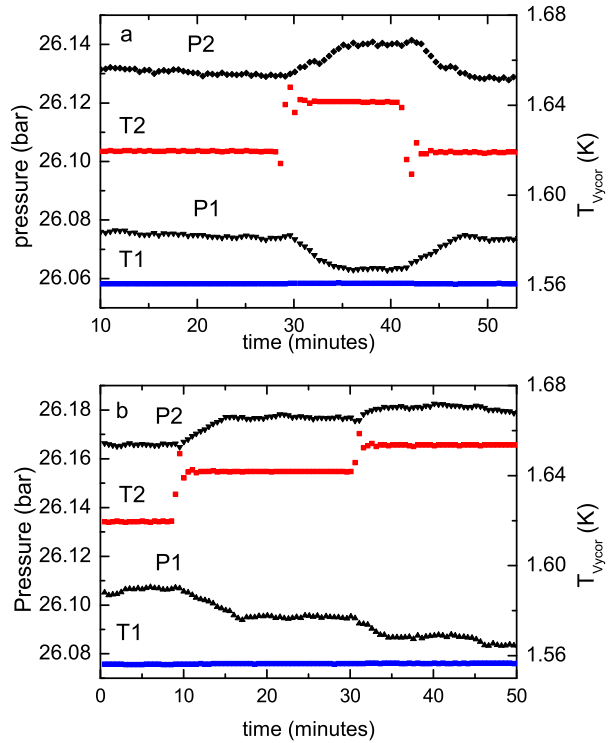}}
\caption{ \label{fig:FS-FT} (color online) (a) Sample FS, $TC$ = 304 mK.  Application of thermal energy to H2, results in a change in $T2$, with $T1$ held constant.  $T2$ is stabilized to its new value in $\approx$ 2 min.  The resulting change in $P2 - P1$ stabilizes in $\approx$ 6 min, with changes in $P1$ and $P2$ linear in time. The rapid small oscillations at the start of changes in T2 result from the electronic temperature stabilization.  (b) Sample FT, $TC$ = 304 mK.  A small overall long-term systematic drift has been removed from the pressure data for sample FT.}
\end{figure}

Figure \ref{fig:FS-FT}a shows data from a solid sample denoted FS at $P$ = 26.1 bar, $TC$ = 304 mK. In the case shown, thermal energy was supplied to H2, which caused $T2$ to change to a larger value while $T1$ was held constant.  $\Delta T$ stabilized within 2 min.  The result, evident in the figure, was an approximately linear increase (decrease) in $P2$ ($P1$) to a new stable value with an increase in $T2$.  For the parameters here, one would expect using eq.1 that at $t \sim $ 20 min $P2 - P1$ = 0.056 bar; 0.052 bar is measured.  In the vicinity of $t \sim$ 35 min, $P2 - P1$ is predicted to be 0.074 bar and 0.075 bar is measured.   The rate of change of, for example, $P2$ following the change in $T2$ near t = 29 min was $dP2/dt \approx$ 2.0 mbar/min with $\Delta P$ stable in $\delta t \approx$ 6 min.  With liquid helium in the apparatus at 250 mK and $P \approx$ 24.4 bar a similar $\Delta T$ created a similar $\Delta P$, but with the result $dP2/dt \approx$ 22 mbar/min., a much faster rate, which indicates clearly that the epoxy-coated Vycor does not restrict the flux; the presence of the solid sample limits the flux.

Another solid sample shown in Figure \ref{fig:FS-FT}b, FT, produced a constant flux similar to sample FS with $dP2/dt \approx$ 2.2 mbar/min.  Two conclusions from samples like FS and FT are apparent: (1) a change in the temperature difference $T2 - T1$ results in a flux of atoms to produce a change in the pressure difference $P2 - P1$ that is in agreement with expectations based on the fountain effect and (2) the change in $P2 - P1$ with time is linear, which suggests that there is a critical flux imposed by the presence of the solid helium, which is a characteristic of the flow of a superfluid limited by a critical velocity.

\begin{figure}
\resizebox{3.5 in}{!}{
\includegraphics{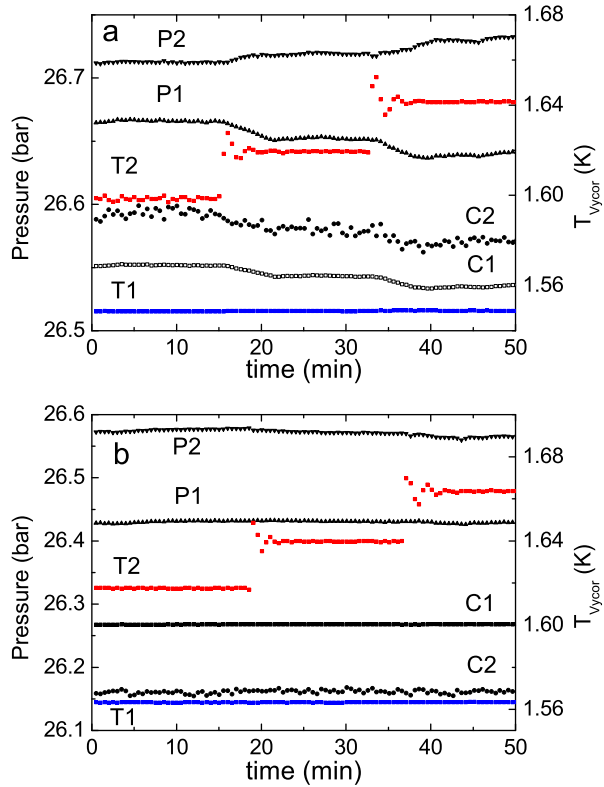}}
\caption{ \label{fig:FIFM} (color online) (a) Sample FI, $TC$ = 250 mK.  Flow was induced through the solid by changing $T2$.  (b) Sample FM, $TC$ = 700 mK.  Changes to $T2$ resulted in no corresponding changes in $P1, P2, C1$ or $C2$.}
\end{figure}

Figure \ref{fig:FIFM}a shows data from sample FI, grown at $P$ = 26.7 bar and studied at $TC = 250$ mK.  Changes in $\Delta T$ for this sample produced changes in $\Delta P$ similar to those seen in samples FS and FT. Also shown here (and seen in samples FS and FT) are changes in $C1$ and $C2$ induced by the changes in $\Delta T$. Note that with $T1$ fixed, a change in $P2$ and $P1$ is induced by a change in $T2$; atoms are supplied to R2 from R1, but also from the solid. This may be a manifestation of the isochoric compressibility of the solid\cite{Soyler2009,Ray2010a}.  Changes in $C1$ and $C2$ maintain the fountain effect along the Vycor between the cell and R1 and R2.  Figure \ref{fig:FIFM}b shows data from sample FM, which was grown and studied at $P$ = 26.5 bar, $TC = 700$ mK.  Here changes to $T2$ produced very small changes in $P2$, and no corresponding change in $P1$.  Also, note that the small changes in $P2$ due to changes in $T2$ are opposite what would be expected for a fountain effect.  These small pressure changes are likely due to a small density change of the helium in the reservoir as the temperature in the reservoir changes.  In fig.~\ref{fig:FIFM}b $T2$ changed from 1.617 K to 1.641 K.  Over this temperature range, the density changes\cite{Maynard1976} by $\Delta \rho / \rho  \approx 8.5 \times 10^{-4}$.  Also, there were no changes in the cell pressure, such as those that were observed in fig.~\ref{fig:FIFM}a.  Thus, at $TC = 700$ mK we were unable to induce a mass flow through the solid region by changing $T1$ or $T2$.  Whatever was conducting the flow in sample FI at 250 mK seen in fig.~\ref{fig:FIFM}a did not do so for sample FM at 700 mK.  This observation is consistent with our previous results \cite{Ray2009b} in which we were unable to induce mass flow across the solid when we changed the pressure in one fill line for $TC \gtrsim 550$mK.  Finally, this sample created at 700 mK was cooled to $TC = 300$ mK, and a change in $\Delta T$ did not induce a change in $\Delta P$ between the two reservoirs.  This is also consistent with our previous observations, which showed that samples created at higher temperatures don't flow when cooled undisturbed to lower temperatures. Fresh samples created at low temperatures allow a mass flow \cite{Ray2009b}.

In addition to the fountain pressure that can be measured between the two reservoirs when a temperature difference is imposed between them, there is a fountain pressure between the helium in the Vycor rods in the cell and the respective individual reservoirs due to the temperature difference between the cell and the reservoirs.  This remains, of course, even for the case when $T1 = T2$.  When we measure the pressure between the cell and the reservoirs, $\delta Pi = Pi - Pcell$ {\it vs.} $\delta Ti = Ti - Tcell, i = 1,2$, we find quantitative agreement with the fountain effect for liquid helium in the cell at 24.34 bar and $TC = 250$ mK for reservoir temperatures we have studied (1.52 K $< T <$ 1.64 K).   With solid in the cell at 250 mK, a given solid sample shows the same dependence of $\delta Pi$ on $\delta Ti$ as that with liquid in the cell, but with a modest sample to sample offset consistent with an inability to precisely know the sample pressure due to the presence of modest stable gradients\cite{Ray2009b,Ray2010a} that are sometimes observed with solid in the cell.

As long as there is a path for helium to flow through the solid any change in the chemical potential in one reservoir will be communicated to the other reservoir via these flow paths until the chemical potentials are equal in both reservoirs.  In other words, if $T1 = T2$ then $P1 = P2$ regardless of the presence of a pressure gradient in the solid lattice that might be recorded by C1 or C2 as long as there are flow paths through the solid.  Thus, the pressure difference between the R1 and R2 induced by $T1 - T2 \neq$ 0 is given by eq.1.

\begin{figure}
\resizebox{3.5 in}{!}{
\includegraphics{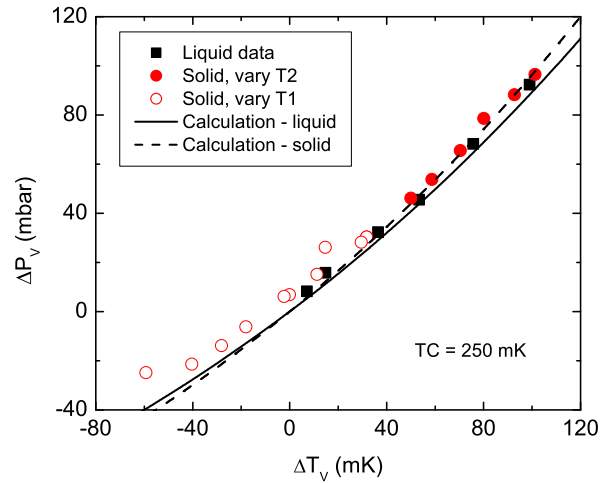}}
\caption{\label{fig:fountain} (Color online) Measured pressure difference between R1 and R2, $\Delta P_V = P2 - P1$ at $TC = 250$ mK for a temperature difference, $\Delta T_V = T2 - T1$ between the reservoirs. The lines show eq. 1 for $P = 24.73$ bar and $P = 26.75$ bar, which is the pressure in the reservoirs for the liquid data and solid data (sample FI) respectively. }
\end{figure}
Figure \ref{fig:fountain} shows the pressure difference between the two liquid reservoirs, $\Delta P_V = P2 - P1$ vs. the temperature difference between the two reservoirs, $\Delta T_V = T2 - T1$ for sample FI at $TC = 250$ mK, $PC \approx 26.58$ bar and for liquid in the cell at $TC = 250$ mK, $PC = 24.34$ bar along with the expectation based on eq. 1 for reservoir pressures $P1,P2 = 24.73$ bar (the liquid data) and for $P1,P2 = 26.75$ bar (solid data).  The data superimpose, and agree well with eq. 1. Data from samples FS and FT show similar, good agreement with eq. 1.

The detailed nature of these flow paths is still not known although there have been several predictions.  For example, grain boundaries\cite{Pollet2007} and dislocations\cite{Boninsegni2007} have been shown in simulations to support mass flow, as has a glassy, or ``superglass" phase of solid \4he \cite{Boninsegni2006,Biroli2008,Andreev2007}.  We have previously reported on quantitative aspects of some of these scenarios, and how they relate to our earlier observations \cite{Ray2008a,Ray2009b}.  Another possibility is the presence of superfluid cores along edge dislocations as proposed by Soyler {\it et al}\cite{Soyler2009}. This proposal explains two aspects of our experiments: (1) the flux of atoms though solid helium below a characteristic temperature and (2) the growth of the density of the solid at constant volume in the presence of a chemical potential gradient.  It has been suggested that the flow might be due to liquid channels which could exist in the solid\cite{Balibar2008a}, but as we have noted previously\cite{Ray2008a,Ray2008e}, this scenario seems to be inconsistent with a number of our observations; in particular, it seems inconsistent with the lack of a mass flow under any circumstances for solids at temperatures above $TC \gtrsim 550$ mK.  While it is not possible for us to identify the specific process that leads to the mass flux that we observe, the presence of a fountain effect with solid helium in the cell and linear changes in $\Delta P$ with an imposed change in $\Delta T$ allows the conclusion that a flux-limited pathway percolates the solid.  It is not possible to identify the precise bottleneck that limits the flow; it could be the pathways in the solid or it could be at the interface between the solid and the Vycor.

Unclear at present is the connection of this work to the numerous experiments that have been heavily focused on the behavior of solid helium in the temperature range below $\approx$ 100 mK.  There may well be no connection.

In summary, we have induced a mass flow through solid \4he by creating a chemical potential difference (by means of the imposition of a temperature difference) between two liquid reservoirs that are connected by Vycor rods in series with solid \4he.  When the temperature in one reservoir is changed, we observe a corresponding change of the pressure, which agrees with predictions based on the fountain effect.  In order for this to happen, mass was supplied through the cell that contained the solid helium sample and this happened with a rate-limited flux.  At 700 mK, equilibration is not achieved, which is consistent with our previous results.  Our observations confirm that the relevant quantity that induces flow in our experiments is the chemical potential.

 We thank N. Prokofev and B. Svistunov for numerous discussions.  This work was supported by NSF DMR  07-57701 and 08-55954,.

\bibliography{ref}% Produces the bibliography via BibTeX.

%Included for Gather Purpose only:
%input "C:\Documents and Settings\mike\localtexmf\bibtex\bib\ref.bib"
\end{document}